\documentclass[aps,prd,nofootinbib]{revtex4}
\pdfoutput=1

\usepackage{amsmath}
\usepackage{amssymb}
\usepackage{anysize}
\usepackage{bold-extra}
\usepackage{color}
\usepackage{enumerate}
\usepackage{fancyhdr}
\usepackage{graphicx}
\usepackage[linktocpage,colorlinks,urlcolor=blue]{hyperref}
\usepackage{mathrsfs}
\usepackage{multirow}
\usepackage[final]{pdfpages}
\usepackage{rotating}
\usepackage{subfig}
\usepackage{textcomp}
\usepackage{url}
\usepackage{verbatim}
\usepackage{wrapfig}
\usepackage{amsfonts}
\usepackage{comment}
\usepackage{epigraph}
\usepackage[utf8]{inputenc}
\usepackage{slashed}
\usepackage{bbm}
\usepackage{xfrac}
\def\nn{\nonumber}

\def\lsim{\mbox{\raisebox{-.6ex}{~$\stackrel{<}{\sim}$~}}}

\begin{document}

\title{NLO Perturbativity Bounds on Quartic Couplings in Renormalizable Theories with $\phi^4$-like Scalar Sectors}

\author{Christopher W. Murphy}
\email{cmurphy@quark.phy.bnl.gov}
\affiliation{Department of Physics, Brookhaven National Laboratory, Upton, N.Y., 11973, U.S.A.}

\begin{abstract}
The apparent breakdown of unitarity in low order perturbation theory is often is used to place bounds on the parameters of a theory.
In this work we give an algorithm for approximately computing the next-to-leading order (NLO) perturbativity bounds on the quartic couplings of a renormalizable theory whose scalar sector is $\phi^4$-like.
By this we mean theories where either there are no cubic scalar interactions, or the cubic couplings are related to the quartic couplings through spontaneous symmetry breaking.
The quantity that tests where perturbation theory breaks down itself can be written as a perturbative series, and having the NLO terms allows one to test how well the series converges.
We also present a simple example to illustrate the effect of considering these bounds at different orders in perturbation theory.
For example, there is a noticeable difference in the viable parameter when the square of the NLO piece is included versus when it is not.
\end{abstract}

\maketitle

%%%%%%%%%%%%%%%%%%%%%%
\section{Introduction}
The unitarity of the $S$-matrix is frequently used to place theoretical constraints on the parameters of a theory.
On the one hand, if a $2 \to 2$ scattering amplitude grows with energy, as is typically the case in non-renormalizable theories, then the condition $S^{\dagger} S = \mathbbm{1}$ will inevitably be violated at some energy scale.
This energy scale then sets an upper limit on where new degrees-of-freedom must appear to unitarize the scattering amplitude. 
While is interesting in its own right, it is not the focus of this work.
On the other hand, in renormalizable theories, there are relations among the parameters of the theory that cancel this growth with energy~\cite{Gunion:1990kf}. 
Nevertheless the same procedure can be used to place ``perturbativity bounds'' on the parameters of a renormalizable theory.
Most famously, a leading order (LO) analysis of this type yielded an upper bound on the mass of the Higgs boson in the Standard Model (SM), $m_h \lsim 1$~TeV~\cite{Dicus:1992vj, Lee:1977yc, Lee:1977eg}. 
If some combination of parameters in a renormalizable theory are too large, the amplitude will appear to be non-unitary at some order in perturbation theory. 
Of course these theories are unitary.
The more accurate statement is that perturbation theory is breaking down for this choice of parameters.

This method has subsequently been improved and refined. 
See for instance~\cite{Marciano:1989ns, Dawson:1988va, Durand:1992wb, Durand:1993vn} for studies of beyond leading order effects in the SM, including both renormalization group (RG) improvement and higher fixed order contributions.
If some choice of parameters violates the perturbativity bound at tree level it may be that perturbativity is restored at one-loop, or perhaps it may be that those parameters are not viable at any (low) order in perturbation theory.
The preceding discussion calls attention to the following fact.
The quantity that tests where perturbation theory breaks down itself can be written as a perturbative series.
If higher order terms are known, one is then able to test how well the series converges. 
To date not much work has been done along these lines in theories beyond the SM. 
The one-loop corrections necessary to compute perturbativity bounds at next-to-leading order (NLO) in the Two-Higgs Doublet Model (2HDM) with a softly broken $\mathbf{Z_2}$ symmetry were computed in Ref.~\cite{Grinstein:2015rtl}. 
Ref.~\cite{Cacchio:2016qyh} then performed a comprehensive analysis of the viable parameter space in the 2HDM using these NLO perturbativity bounds. 
Prior to this work, an \textit{ansatz} inspired by SM results was used to to estimate the higher order corrections in the 2HDM~\cite{Baglio:2014nea, Chowdhury:2015yja}. 
Not only did the work of Ref.s~\cite{Grinstein:2015rtl, Cacchio:2016qyh} resolve the ambiguity of how to implement perturbativity bounds at NLO in the 2HDM, but they also revealed where the dominate NLO contribution comes from in the SM and the 2HDM.

In this work we first derive a formula for the functional form of the partial wave matrix for high energy $2 \to 2$ scalar scattering in a general renormalizable theory.
This allows us to construct an algorithm for approximately computing the NLO perturbativity bounds on the scalar quartic couplings of a general renormalizable theory, (approximately) generalizing the result of~\cite{Grinstein:2015rtl, Cacchio:2016qyh}.
Ref.~\cite{Grinstein:2015rtl} showed that this approximation dominates the NLO contribution to the perturbativity bounds in the SM and certain special cases of the 2HDM.
Ref.~\cite{Cacchio:2016qyh} went further and showed that this is generally a good approximation in the 2HDM with a (softly broken) $\mathbf{Z_2}$ symmetry.
Since the approximation is based on the pattern found in~\cite{Grinstein:2015rtl, Cacchio:2016qyh} we expect that this approximation should generally be a good estimate of the full NLO contribution in theories whose scalar sector is $\phi^4$-like. 
By this we mean theories where either there are no cubic scalar interactions, or the cubic couplings are related to the quartic couplings through spontaneous symmetry breaking.
 
This approximate NLO result only requires knowledge of the leading order matrix of partial wave amplitudes, and the one-loop scalar contribution to the beta function of each quartic coupling that is to be bounded. 
The advantage of this approximation is its simplicity as both of the required quantities are relatively easy to determine.
To further ease the calculation of these NLO bounds a package implementing this algorithm, $\mathtt{NLOUnitarityBounds}$, is available at~\url{https://github.com/christopher-w-murphy/NLOUnitarityBounds} in both Mathematica and Jupyter Notebook formats.
%\href{https://github.com/christopher-w-murphy/NLOUnitarityBounds}{this https URL}.
A (likely incomplete) list of other models for which all of the results necessary to implement this algorithm are already known is: the Manohar-Wise Model~\cite{He:2013tla}, its extension to include an additional color singlet, $SU(2)_L$ doublet~\cite{Cheng:2016tlc, Cheng:2017tbn}, and the Left-Right Symmetric Model~\cite{Chakrabortty:2016wkl}.

As we will show, the natural interpretation of the couplings appearing in the NLO partial wave amplitudes are RG improved couplings evaluated at a scale much larger than the typical scales of the theory.
As such, these corrections are naturally useful in analyses at high scales. 
Examples of this include high scale flavor alignment~\cite{Gori:2017qwg}, effects of custodial symmetry breaking at high energies~\cite{Blasi:2017xmc}, or simply investigating the validity of a model up to a high energy scale~\cite{Grinstein:2013npa, Chakrabarty:2014aya, Chowdhury:2015yja, Ferreira:2015rha, Cheng:2017tbn}.

The rest of this paper starts with a brief review of perturbative unitarity in Sec.~\ref{sec:rev}. 
Following that is a derivation of our main results in Sec.~\ref{sec:derv}. 
We then present an example to illustrate the effect of considering perturbativity bounds at different orders in perturbation theory in Sec.~\ref{sec:ex}.
The example is simple, but it serves to highlight how the viable parameter space in a model can change from order-to-order in perturbation theory.
In particular, there is a noticeable difference in the viable parameter when the square of the NLO piece is included versus when it is not.
The implementation of this model is included in the example notebook associated with the $\mathtt{NLOUnitarityBounds}$ package.
Lastly, we discuss our findings in Sec.~\ref{sec:sum}.

%%%%%%%%%%%%%%%%%%%%%%
\section{Brief Review of Perturbative Unitarity}
\label{sec:rev}
In this section we give a brief review of perturbative unitarity.  
The $S$-matrix is unitary, $S^{\dagger} S = \mathbbm{1}$. 
This condition can be translated into a relation among the various partial wave amplitudes of a given theory, see \textit{e.g.}~\cite{Durand:1992wb, Durand:1993vn, Itzykson:1980rh, Kilian:2014zja}
\begin{equation}
\label{eq:smat}
\text{Im}\left(\mathbf{a}_j^{2 \to 2}\right) = \left(\mathbf{a}_j^{2 \to 2}\right)^{\dagger} \mathbf{a}_j^{2 \to 2} + \sum_{n > 2} \left(\mathbf{a}_j^{2 \to n}\right)^{\dagger} \mathbf{a}_j^{2 \to n} ,
\end{equation}
where $\mathbf{a}_j^{2 \to 2}$ is computed in a basis such that it is diagonal, \textit{i.e.}~it is the eigenvalues of $\mathbf{a}_j^{2 \to 2}$ that satisfy the relation~\eqref{eq:smat}. 
Note also that an integral over the $n$-body phase space for each term in the sum in the rightmost term in~\eqref{eq:smat} is left understood. 
This is nothing but the equation for an $n$-sphere of radius $\tfrac{1}{2}$ centered at $\text{Re}(\mathbf{a}_j^{2 \to 2}) = |\mathbf{a}_j^{2 \to n}| = 0$, $\text{Im}(\mathbf{a}_j^{2 \to 2}) = \tfrac{1}{2}$. 

The imaginary part of $\mathbf{a}_j^{2 \to 2}$ is fixed by the real part of $\mathbf{a}_j^{2 \to 2}$ and the $2 \to n$ partial wave amplitudes. 
Focusing on the eigenvalues, $a_j^{2 \to 2}$, of the matrix $\mathbf{a}_j^{2 \to 2}$ we have
\begin{equation}
\label{eq:Im}
2\, \text{Im}\left(a_j^{2 \to 2}\right)_{\mp} = 1 \mp \sqrt{1 - 4 \mathcal{A}_j^2} , \quad \mathcal{A}_j^2 = \left[\text{Re}\left(a_j^{2 \to 2}\right)\right]^2 + \sum_{n > 2} \left| a_j^{2 \to n}\right|^2 ,
\end{equation}
with the $- (+)$ solution corresponding to the case when the imaginary part of $a_j^{2 \to 2}$ is less than (greater than) one-half. 
Assuming a perturbative expansion is viable, the first few orders of Eq.~\eqref{eq:Im} take the form
\begin{align}
\label{eq:Imexp}
\text{Im}\left(a_j^{(0)}\right)_{-} & = 0 , \\
\text{Im}\left(a_j^{(1)}\right)_{-} & = \left(a_j^{(0)}\right)^2 , \nn \\
\text{Im}\left(a_j^{(2)}\right)_{-} & = 2 a_j^{(0)}\, \text{Re}\left(a_j^{(1)}\right) + \left|a_j^{2 \to 3, (0)}\right|^2 , \nn
\end{align}
where the superscript $(\ell)$ is the perturbative order and we have dropped the superscript $2 \to 2$. 
For the $+$ solution we instead have $\text{Im}(a_j^{(0)})_{+} = 1$, $\text{Im}(a_j^{(\ell > 0)})_{+} = - \text{Im}(a_j^{(\ell)})_{-}$. 
Assuming a valid perturbative expansion, the $+$ solution corresponds to a scattering amplitude whose tree level imaginary part is $\text{Im}(\mathcal{M}^{(0)}) = 16 \pi$. 
Since this is not a frequently encountered scenario we will not consider this possibility further in this work, drop the subscript $-$, and set an upper limit of $\text{Im}(a_j) \leq \tfrac{1}{2}$ in the process.

Perturbative unitarity bounds are inequalities derived from Eq.~\eqref{eq:smat}. 
From~\eqref{eq:Im} it is clear that the conditions $0 \leq \, \text{Im}(a_j^{2 \to 2}) \leq \tfrac{1}{2}$ and $0 \leq \mathcal{A}_j^2 \leq \tfrac{1}{4}$ yield equivalent bounds when the imaginary part of $a_j^{2 \to 2}$ is kept in its exact form. 
However this is not the case if $\text{Im}(a_j^{2 \to 2})$ is expanded to a finite order in perturbation theory, as in~\eqref{eq:Imexp}. 
For example, to two-loop order, the corresponding bounds are always weaker than those from $0 \leq \mathcal{A}_j^2 \leq \tfrac{1}{4}$. 
Explicitly, to leading order, $0 \leq \, \text{Im}(a_j^{2 \to 2}) \approx (a_j^{(0)})^2 \leq \tfrac{1}{2}$ whereas $0 \leq \mathcal{A}_j^2 \approx (a_j^{(0)})^2 \leq \tfrac{1}{4}$. 
Starting at three-loops in $\text{Im}\left(a_j\right)$, NNLO for $\mathcal{A}_j$, the relative strength of the two bounds in no longer fixed.

Once the approximate NLO contributions to the eigenvalues of the partial wave matrix 
 are known, which are derived in Sec.~\ref{sec:derv}, perturbative unitarity bounds can be obtained by evaluating one or more of the following
\begin{align}
\label{eq:Ubounds}
&\text{LO:} \quad \left(a_0^{(0)}\right)^2 \leq \frac{1}{4}, \\
&\text{NLO:} \quad 0 \leq \left(a_0^{(0)}\right)^2  + 2 \left(a_0^{(0)}\right) \text{Re}\left(a_0^{(1)}\right) \leq \frac{1}{4} , \nn \\
&\text{NLO+:} \quad \left[\left(a_0^{(0)}\right) + \text{Re}\left(a_0^{(1)}\right)\right]^2 \leq \frac{1}{4}. \nn
\end{align}
If perturbation theory is valid it is expected that the bounds obtained from the upper limit ($\leq \tfrac{1}{4}$) will be similar in all three cases since the LO eigenvalue is non-trivial and the NLO(+) piece should represent a small correction, but it is important to test, and confirm or deny, if this is actually the case.
The  bound obtained from the lower limit ($0 \leq$) originates from Ref.~\cite{Durand:1992wb}, Eq.~(75) in particular.\footnote{This perturbativity bound was subsequently investigated in the 2HDM in Ref.~\cite{Grinstein:2015rtl} where it was called $R_1$ (not $R_1^{\prime}$).} 
In contrast to the $\leq \tfrac{1}{4}$ bound, the $0 \leq$ bound only becomes non-trivial at NLO, and can only be violated when perturbation theory breaks down.
While it is expected that the parameter space of a given theory will be more constrained by the NLO perturbativity bounds because of the additional $0 \leq$ handle, using this bound goes somewhat against the spirit of the introduction in that there is no similar higher-order term to compare with.
In fact, the NLO+ expression shows how the apparent $0 \leq$ violation of unitarity is resolved at higher-orders in perturbation theory. 
(Note that the NLO+ expression contains some NNLO terms, but of course is not the full NNLO expression.) 
To bring things full circle, at NNLO similar apparent $0 \leq$ violations of unitarity can occur from the interference between the tree level and two-loop $2 \to 2$ amplitudes and/or the tree level and one-loop $2 \to 3$ amplitudes.  
However, there is no reason to expect the higher-order versions of the $0 \leq$ bound to be similar to the NLO version, as was the case for the $\leq \tfrac{1}{4}$ bound, since there is little and/or no overlap between the potentially problematic terms.

%%%%%%%%%%%%%%%%%%%%%%
\section{Generic Partial Wave Amplitudes for High Energy Scalar Scattering}
\label{sec:derv}
\subsection{Elements of the Partial Wave Matrix}
Consider a potential of the form
\begin{equation}
V = \tfrac{1}{2} m_{\alpha}^2 \phi_{\alpha} \phi_{\alpha} + \kappa_{\alpha \beta \gamma} \phi_{\alpha} \phi_{\beta} \phi_{\gamma} + \lambda_{\alpha \beta \gamma \delta} \phi_{\alpha} \phi_{\beta} \phi_{\gamma} \phi_{\delta},
\end{equation}
where the subscript Greek letters are flavor indices.
The $2 \to 2$ scattering of high energy ($E \gg |\kappa_{\alpha \beta \gamma}|, m_{\alpha}$) scalars can schematically by written as
\begin{equation}
\mathcal{M}_{i \to f} = \left(Z_{\phi}^{1/2}\right)^4 \mathbf{V}\left[\phi^4\right] ,
\end{equation}
where $(Z_{\phi}^{1/2})^4$ is the product of the four external wavefunction renormalization factors, and $\mathbf{V}\left[\phi^4\right]$ is the four-point function.
At tree level the unrenormalized four-point function is simply a linear combination of quartic couplings,
\begin{equation}
\label{eq:tree}
\mathbf{V}\left[\phi^4\right]_{\text{tree}} = - c_m \lambda_{m B} = - c_m \left(\lambda_m + \delta \lambda_m\right) ,
\end{equation}
where $\delta \lambda_m$ is the counterterm associated with the renormalized coupling $\lambda_m$, $c_m$ is the numeric coefficient associated with a given $\lambda_m$, and a subscript Roman letter is shorthand for a set of Greek letters, \textit{e.g.} $m = \alpha \beta \gamma \delta$.
At high energies the diagrams involving cubic couplings generally do not contribute to Eq.~\eqref{eq:tree}.
For $s$-channel processes this is manifestly true simply because we assume $E \gg |\kappa_{\alpha \beta \gamma}|$.
To see this is also the case for $t$- and $u$-channel processes one should retain the full mass dependence of the diagram until after the partial wave amplitude is computed at which point it is safe to take the high energy limit.
An exception to this occurs when all of the particles, internal and external, in a digram are massless.
Such a situation would occur if the theory contains a neutral, $CP$-even Goldstone boson since it would then be possible to write down vertices with an odd number of this Goldstone boson.
In this case there is a physical divergence in the forward region, analogous to Rutherford scattering, and this method as it is currently implemented is not applicable.
However it is worth pointing that a careful study of the analytic structure of amplitudes involving the $t$-channel exchange of massless particles showed that the na\"{i}ve sum rules for processes such as $W_L^+ W_L^-$ scattering are still correct~\cite{Bellazzini:2014waa}, so perhaps there may still be a way to extract perturbativity bounds from such amplitudes. 
In any case, henceforth we will neglect the trilinear couplings $\kappa_{\alpha \beta \gamma}$.

A generic one-loop diagram in $D = 4 - 2 \varepsilon$ dimensions with four external and two internal scalars, which is the only topology of all scalar, 1PI diagrams that persists in the high energy limit, takes the following form 
\begin{equation}
\frac{\lambda_m \lambda_n}{16 \pi^2}\left(\frac{1}{\varepsilon} + 2 - \ln\left(\frac{- p^2 - i 0_{+}}{\mu^2}\right)\right) .
\end{equation}
As is typically done the scale $\mu$ has been introduced to keep the quartic couplings dimensionless. 
The sum of all such diagrams leads to the four-point function at the one-loop level
\begin{align}
\mathbf{V}\left[\phi^4\right] &= - c_m \lambda_m + \frac{\lambda_m \lambda_n}{16 \pi^2} \left[\left(\sigma_{mn} + \tau_{mn} + \upsilon_{mn}\right) \left(\frac{1}{\varepsilon} + 2 + \ln\left(\frac{s}{\mu^2}\right)\right) \right. \\
&\left. + i \pi \sigma_{mn}  - \tau_{mn} \ln\left(\frac{- t}{s}\right) - \upsilon_{mn} \ln\left(\frac{- u}{s}\right) \right] , \nn
\end{align}
where $s$, $t$, and $u$ are the usual Mandelstam variables, and the branch cut in the logarithm yields $\ln(- p^2 - i 0_{+}) \to \ln(p^2) - i \pi$ for $p^2 > 0$. 
The one-loop correction is bilinear in the various couplings with the (model dependent) coefficients $\sigma$, $\tau$, and $\upsilon$ parameterizing the $s$-, $t$-, and $u$-channel contributions, respectively. 

At one-loop the scalar wavefunction renormalization is finite, which allows the beta function of a quartic coupling to be defined simply as
\begin{equation}
\beta_{\lambda_i} = \mu \frac{\partial \mathbf{V}\left[\phi^4\right]}{\partial \mu} ,
\end{equation}
where the particular four-point function entering the definition of the beta function is such that $\mathbf{V}\left[\phi^4\right]_{\text{tree}} = - \lambda_i$. 
From this we see that 
\begin{equation}
\label{eq:beta}
c_m \beta_{\lambda_m} = \left(\sigma_{mn} + \tau_{mn} + \upsilon_{mn}\right) \frac{\lambda_m \lambda_n}{8 \pi^2} ,
\end{equation}
which determines the purely scalar contribution to the one-loop beta-function.

After renormalization the scattering amplitude takes the form
\begin{align}
\mathcal{M}_{i \to f} &= - c_m \lambda_m - c_m \left(\delta \lambda_m\right)_{\text{fin.}} - c_m \lambda_n \left(\delta Z_{mn}\right)_{\text{fin.}} + c_m \beta_{\lambda_m} \left[1 + \ln\left(\frac{\sqrt{s}}{\mu}\right)\right]  \\
&- \frac{\lambda_m \lambda_n}{16 \pi^2} \left[- i \pi \sigma_{mn} + \tau_{mn} \ln\left(\frac{- t}{s}\right) + \upsilon_{mn} \ln\left(\frac{- u}{s}\right) \right] . \nn
\end{align}
The counterterms cancel the divergences arising from the one-loop diagrams, and generically contain finite parts that contribute to the scattering amplitude. 
The diagonal and off-diagonal wavefunction renormalization constants are $\delta Z_{mm}$ and $\delta Z_{mn}$, respectively, both of which are real (except when using a complex-mass scheme).\footnote{In the SM there is a relation between the tree level $2 \to 3$ partial wave amplitudes and the wavefunction renormalization contribution to the one-loop $2 \to 2$ partial wave amplitudes~\cite{Durand:1993vn}, which in our notation takes the form
\begin{equation}
\left|a_j^{2 \to 3, (0)}\right|^2 = \delta Z_{mm} \left|a_j^{2 \to 2, (0)}\right|^2 \quad \text{(in the SM).}
\end{equation}
This causes a partial cancellation in the NLO expression for $\mathcal{A}_j$ that makes our approximation, discussed after~\eqref{eq:elem},  a more accurate representation of $\mathcal{A}_j$, again at least the in SM. It would be interesting to see if (a generalization of) this relation is true in other theories.} 
The off-diagonal terms generally involve a different linear combination of the tree level amplitudes than the diagonal terms. 
The wavefunction counterterms depend on the particular process under consideration, whereas the $\delta \lambda_m$ are process independent.

The full energy dependence of $\mathcal{M}$ can be subsumed into a running coupling using standard renormalization group methods
\begin{equation}
\mu \frac{\partial \bar{\lambda}_m\left(\mu\right)}{\partial \mu} = \beta_{\lambda_m}, \quad \bar{\lambda}_m\left(\mu_{\text{match.}}\right) = \left(\lambda_m\right)_{\text{phys.}} ,
\end{equation}
with $\left(\lambda_m\right)_{\text{phys.}}$ being the combination of physical parameters that defines $\lambda_m$ at the scale $\mu_{\text{match.}}$.
The scattering amplitude now takes the form
\begin{align}
\label{eq:amprun}
\mathcal{M}_{i \to f} &= - c_m \bar{\lambda}_m - c_m \left(\delta \lambda_m\right)_{\text{fin.}} - c_m \bar{\lambda}_n \left(\delta Z_{mn}\right)_{\text{fin.}} + c_m \beta_{\bar{\lambda}_m}  \\
&+ \frac{\bar{\lambda}_m \bar{\lambda}_n}{16 \pi^2} \left[i \pi \sigma_{mn} - \tau_{mn} \ln\left(\frac{- t}{s}\right) - \upsilon_{mn} \ln\left(\frac{- u}{s}\right) \right] . \nn
\end{align}
Retaining only the first term on the right-hand side of Eq.~\eqref{eq:amprun} corresponds to the leading-log (LL) approximation of the NLO contribution.

An element of $\mathbf{a}_j^{2 \to 2}$ is related to the scattering amplitude for the process $\mathcal{M}_{i \to f}$ as follows
\begin{equation}
\label{eq:par}
\left(\mathbf{a}_j^{2 \to 2}\right)_{i, f} = \frac{1}{16 \pi s}\int_{-s}^0 \! dt \, \mathcal{M}_{i \to f}\left(s, t\right) P_j\left(1 + \tfrac{2 t}{s}\right),
\end{equation}
where $P_j$ are the Legendre polynomials. 
In the high energy limit $\mathcal{M}$ is independent of $s$ and $t$ at leading order, allowing us to concentrate on the $j = 0$ case.
Plugging~\eqref{eq:amprun} into~\eqref{eq:par} and simplifying the result using~\eqref{eq:beta} we find
\begin{equation}
\label{eq:elem}
16 \pi \left(\mathbf{a}_0^{2 \to 2}\right)_{i,f} = - c_m \bar{\lambda}_m + \frac{3}{2} c_m \beta_{\bar{\lambda}_m} + \frac{i \pi - 1}{16 \pi^2} \sigma_{mn} \bar{\lambda}_m \bar{\lambda}_n - c_m \left(\delta \lambda_m\right)_{\text{fin.}} - c_m \bar{\lambda}_n \left(\delta Z_{mn}\right)_{\text{fin.}}  .
\end{equation}
Eq.~\eqref{eq:elem} is the exact expression for the functional form of the partial wave matrix of $2 \to 2$ scattering amplitudes in a general renormalizable theory in the high energy limit and assuming the scalar quartic couplings are parametrically larger than the gauge and Yukawa couplings.
To the best of our knowledge this result has not been previously been given in the literature.

%%%%%%
\subsection{Eigenvalues of the Partial Wave Matrix}
In this subsection we give an approximate formula for the NLO corrections to the eigenvalues of the partial wave matrix for high energy scalar scattering. 
One only needs to knows the leading order partial wave matrix and the (scalar one-loop contribution to) the beta functions of the theory under consideration to make use of this approximation.

Recall the well known formula for the NLO perturbations of the eigenvalues of an eigensystem that is known completely at LO
\begin{equation}
\label{eq:EVper}
a_0^{(1)} = \vec{x}^{\top}_{(0)} \cdot \mathbf{a}_0^{(1)} \cdot \vec{x}_{(0)} .
\end{equation}
It says that the NLO eigenvalues depend only on the NLO correction to the matrix and the LO eigenvectors, which are determined from the LO matrix. 
Here the exact and leading order eigensystems respectively are
\begin{align}
\label{eq:exLO}
\mathbf{a}_0 \cdot \vec{x} &= a_0 \, \vec{x} , \\
\mathbf{a}_0^{(0)} \cdot \vec{x}_{(0)} &= a_0^{(0)} \, \vec{x}_{(0)} , \nn
\end{align}
and each object appearing in the first line of~\eqref{eq:exLO} is assumed to have an expansion
\begin{align}
\mathbf{a}_0 &= \mathbf{a}_0^{(0)} + \mathbf{a}_0^{(1)} + \ldots , \\
\vec{x} &= \vec{x}_{(0)} + \vec{x}_{(1)} + \ldots , \nn \\
a_0 &= a_0^{(0)} + a_0^{(1)} + \ldots . \nn
\end{align}

The second term on the right-hand side of Eq.~\eqref{eq:elem} (the $\beta$-function contribution) to an element of $\mathbf{a}_0$ is universal. 
Therefore the $\beta$-function contributions to the eigenvalues of $\mathbf{a}_0$ can simply be determined using Eq~\eqref{eq:EVper}. 
All one has to do to obtain $\mathbf{a}_{0,\beta}^{(1)}$ from $\mathbf{a}_0^{(0)}$ is replace each quartic coupling $\lambda$ that appears in $\mathbf{a}_0^{(0)}$ with $- \tfrac{3}{2} \beta_{\lambda}$. 
Then using the LO eigenvalues $\vec{x}_{(0)}$, which are also determined from $\mathbf{a}_0^{(0)}$, we find the $\beta$ contribution to $a_0^{(1)}$
\begin{align}
a_{0, \beta}^{(1)} &=  \vec{x}_{(0)}^{\top} \cdot \mathbf{a}_{0, \beta}^{(1)} \cdot \vec{x}_{(0)} , \\
\mathbf{a}_{0,\beta}^{(1)} &= - \frac{3}{2} \left. \mathbf{a}_0^{(0)}\right|_{\lambda_m \to \beta_{\lambda_m}} . \nn
\end{align}

Unlike the $\beta$-function contribution, the third term on the right-hand side of Eq.~\eqref{eq:elem} (the $\sigma$ contribution) is not universal.
Without doing an explicit calculation, the $\sigma$ term contribution to the partial wave \textit{matrix} is not known.
However we are interested in knowing the eigenvalues of the matrix rather than the entire matrix itself.
Knowing that the imaginary parts of the NLO elements of $\mathbf{a}_0$ come exclusively from the $\sigma$ contribution, we can use the second line of~\eqref{eq:Imexp}, \textit{i.e.} the fact that the theory is unitary, to determine the $\sigma$ contribution to the NLO eigenvalues. 
Note that the numerical factor in front of $\sigma_{mn}$ in Eq.~\eqref{eq:elem} is complex.
This is convenient as it allows us to use the unitarity of the theory to also determine the real part of the sigma contribution the NLO eigenvalues.
This yields the final result for the sigma contribution
\begin{equation}
a_{0, \sigma}^{(1)} =  \left(i - \frac{1}{\pi}\right) \left(a_0^{(0)}\right)^2 .
\end{equation}  

To proceed further without doing an explicit calculation we must drop the finite pieces of the counterterms.
Thus combining the $\beta$ contribution and the $\sigma$ contribution we arrive at our result for the approximation NLO contribution to the eigenvalues of the partial wave matrix
\begin{equation}
a_0^{(1)} = a_{0,\beta}^{(1)} + a_{0,\sigma}^{(1)} .
\end{equation}

%%%%%%%%%%%%%%%%%%%%%%
\section{A Simple Example}
\label{sec:ex}
In this section we present a simple example to sketch the effect of using perturbativity bounds at different orders in perturbation theory.
Consider as a toy model the 2HDM with a $U(2)$ symmetry, instead of $\mathbf{Z_2}$, to prevent tree level flavor changing neutral currents~\cite{Ivanov:2006yq, Ferreira:2009wh, Branco:2011iw}.
The scalar sector of this model has only two unique quartic couplings, and the $U(2)$ symmetry preserves this relation along the RG-flow.
The potential is
\begin{equation}
V = \frac{\lambda_1}{2} \left[\left(H_1^{\dagger} H_1\right) + \left(H_2^{\dagger} H_2\right) - \frac{v^2}{2}\right]^2  + \left(\lambda_1 - \lambda_3\right) \left[\left(H_1^{\dagger} H_2\right) \left(H_2^{\dagger} H_1\right) - \left(H_1^{\dagger} H_1\right) \left(H_2^{\dagger} H_2\right)\right],
\end{equation}
where $H_{1,2}$ are the two Higgs doublets.\footnote{One could simplify the potential in this case by defining $\lambda_4 \equiv \lambda_1 - \lambda_3$, but we stick with $\lambda_1$ and $\lambda_3$ as they are what is used in~\cite{Branco:2011iw}.}
We will take $\lambda_{1,3}$ to be free parameters for the purposes of this demonstration, and neglect the electroweak vev, $v \approx 246$~GeV, as it is not important at high energies.

\begin{figure}
  \centering
 \subfloat[$\tfrac{1}{4} \geq (a_0^{(0)})^2$]{\includegraphics[width=0.48\textwidth]{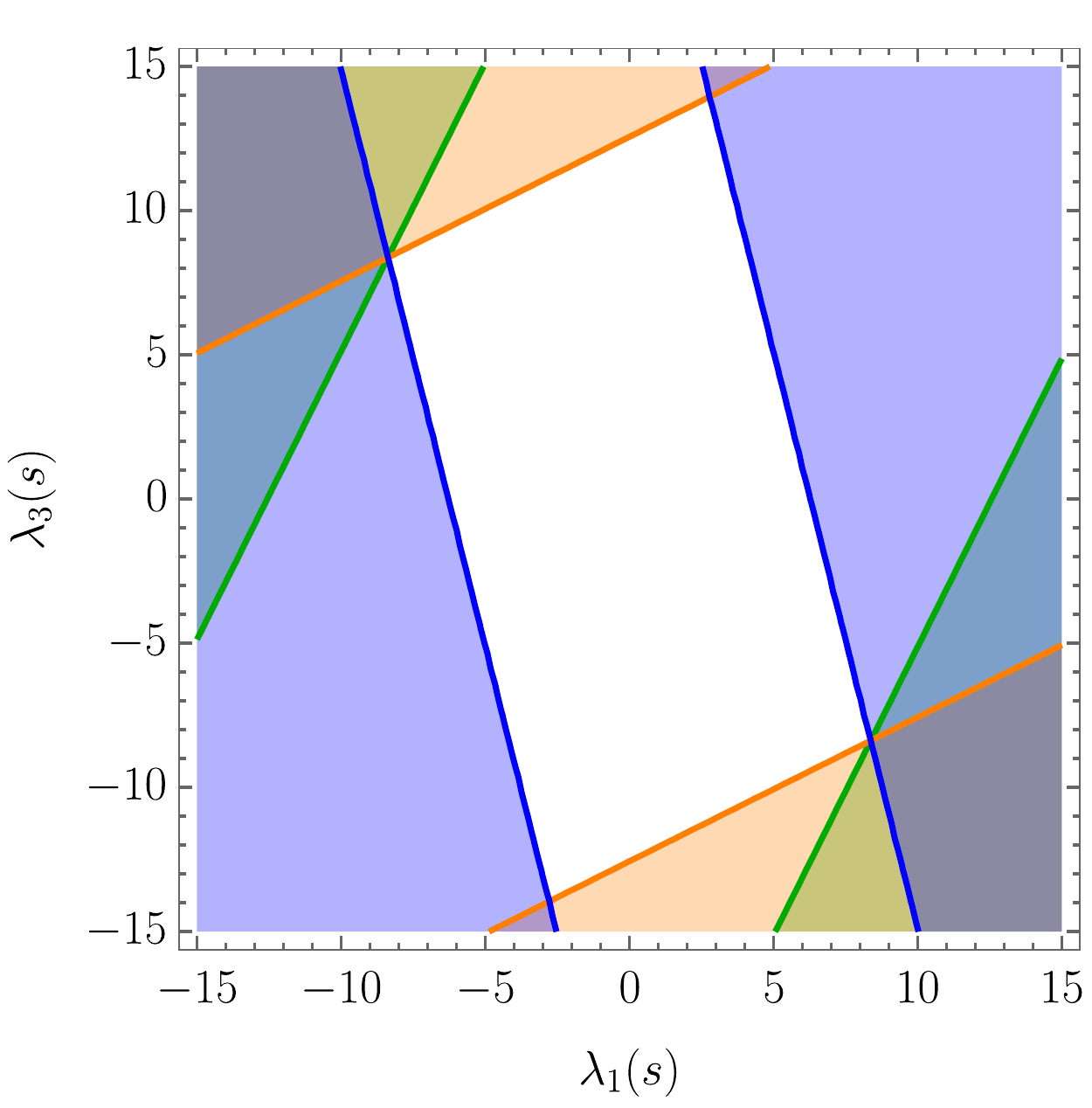}}\,
\subfloat[$\tfrac{1}{4} \geq (a_0^{(0)})^2  + 2 a_0^{(0)} \text{Re}(a_0^{(1)})$]{\includegraphics[width=0.48\textwidth]{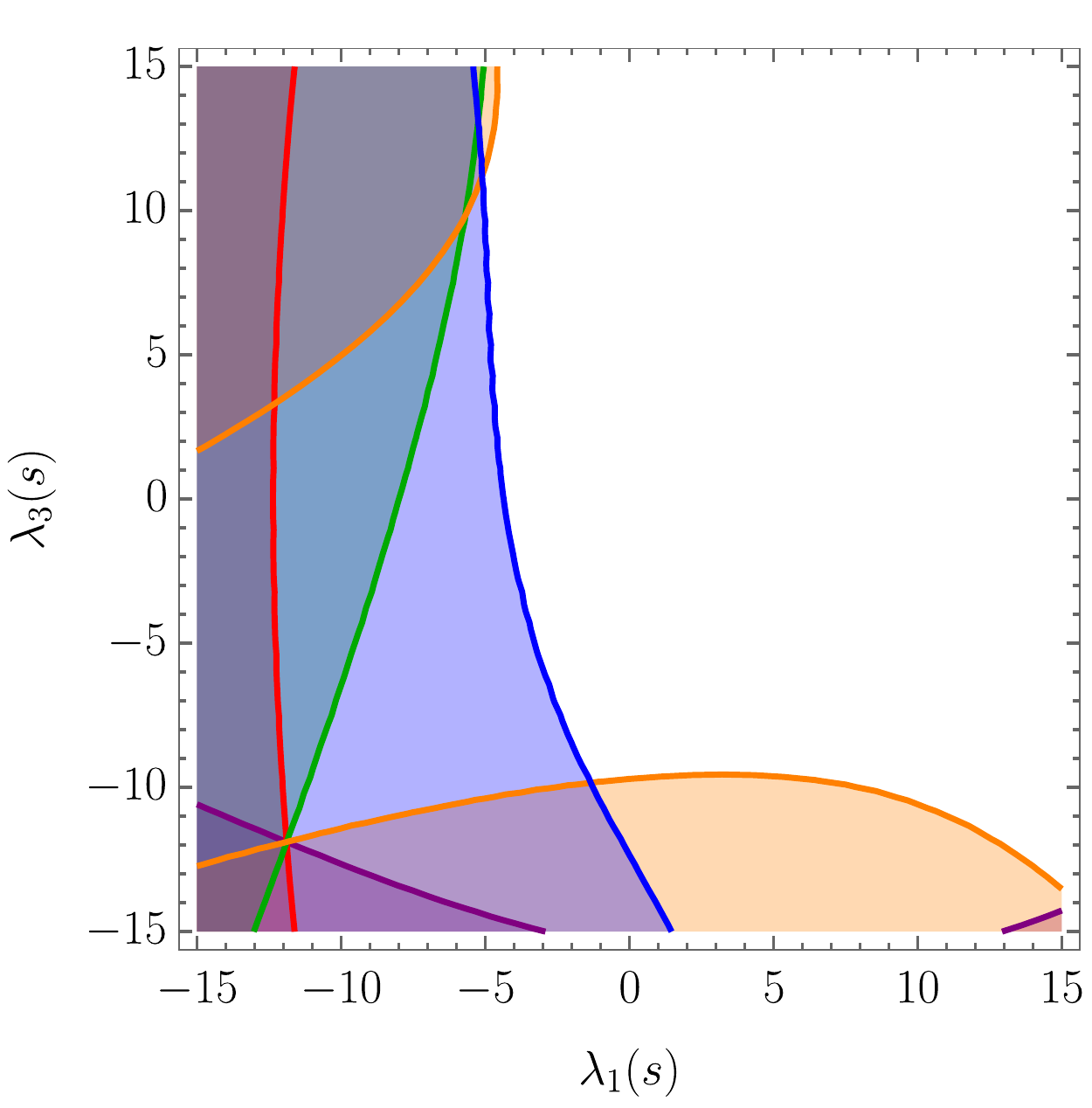}} \\
 \subfloat[$\tfrac{1}{4} \geq (a_0^{(0)} + \text{Re}(a_0^{(1)}))^2$]{\includegraphics[width=0.48\textwidth]{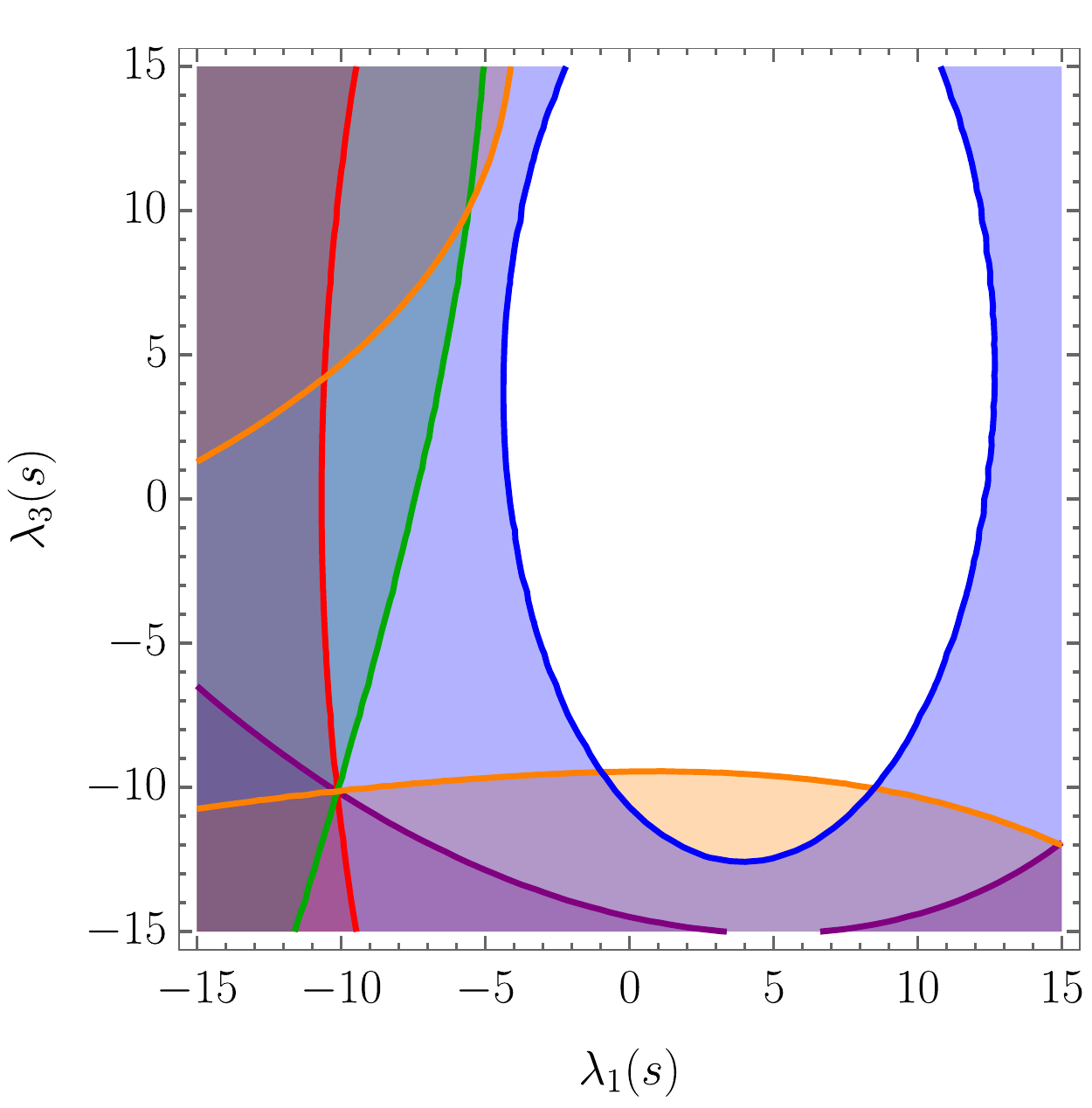}}\,
\subfloat[$0 \leq (a_0^{(0)})^2  + 2 a_0^{(0)} \text{Re}(a_0^{(1)})$]{\includegraphics[width=0.48\textwidth]{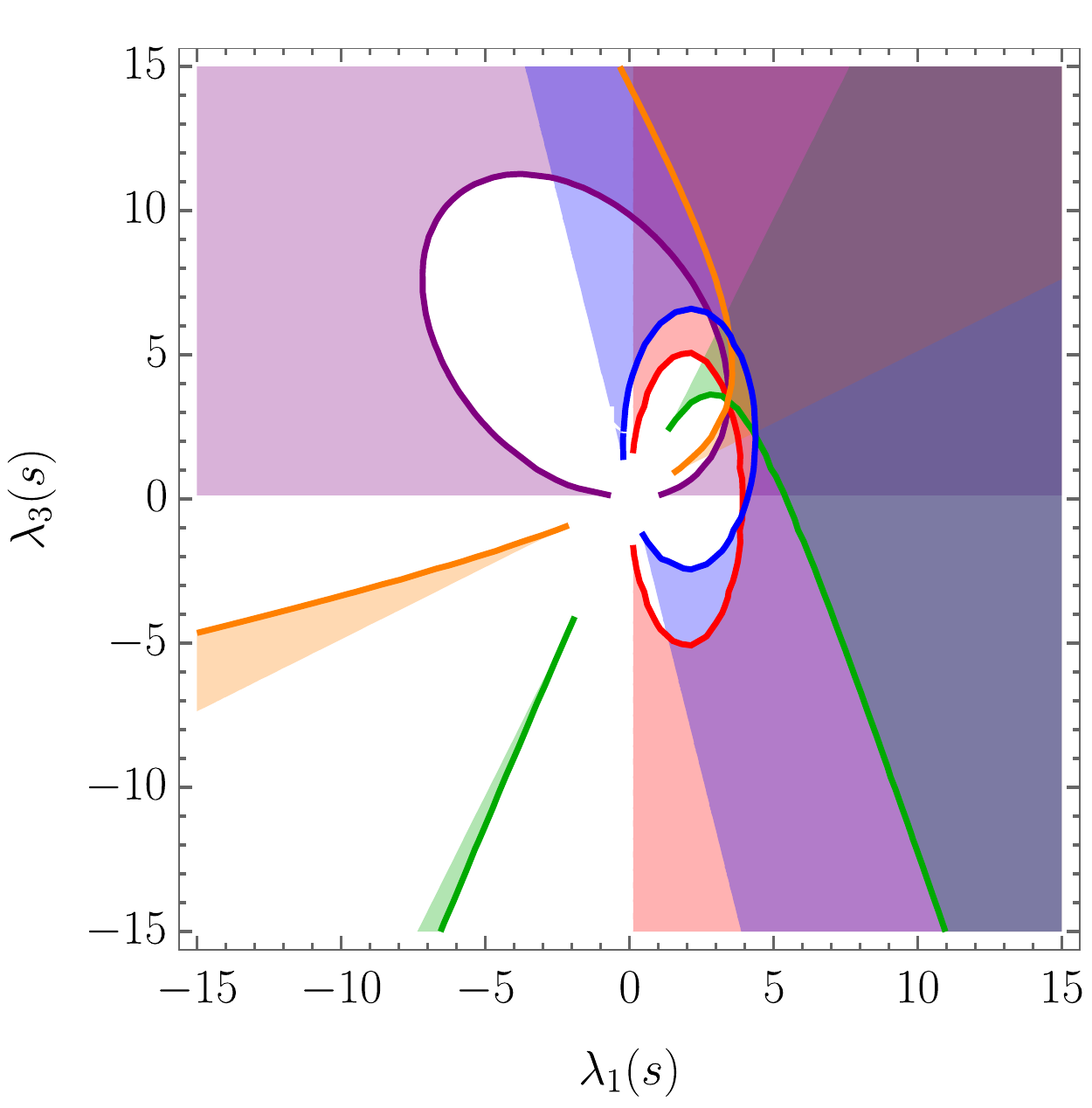}}
 \caption{The bounds on $\lambda_1$ and $\lambda_3$ that result when the various perturbativity conditions are applied with the white space being allowed by all five eigenvalues.}
  \label{fig:U2c}
\end{figure}

\begin{figure}
  \centering
 \subfloat[$- 16 \pi a_0^{(0)} = 4 \lambda_1 + \lambda_3 $]{\includegraphics[width=0.48\textwidth]{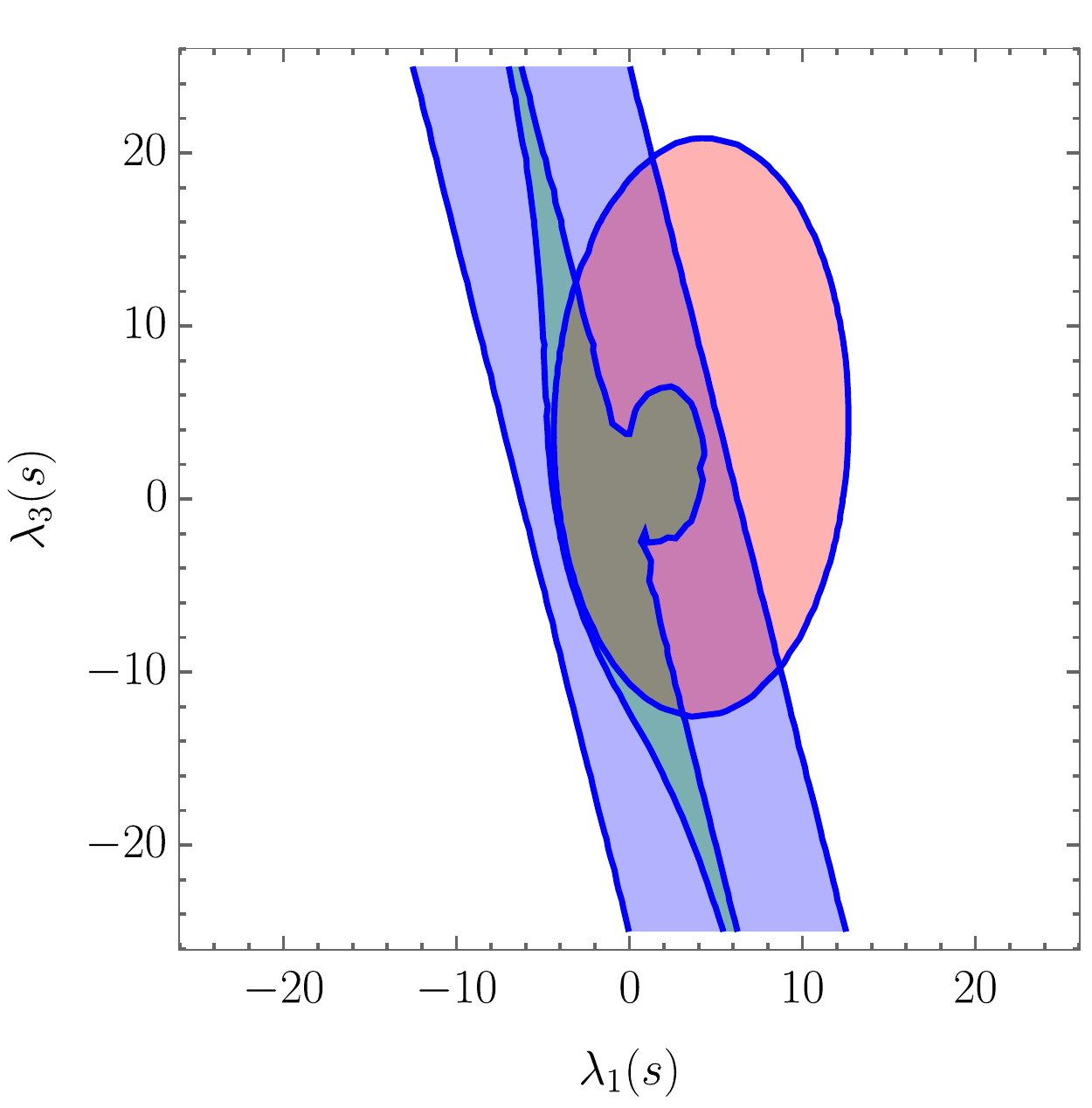}}\,
\subfloat[$- 16 \pi a_0^{(0)} = - \lambda_1 + 2 \lambda_3 $]{\includegraphics[width=0.48\textwidth]{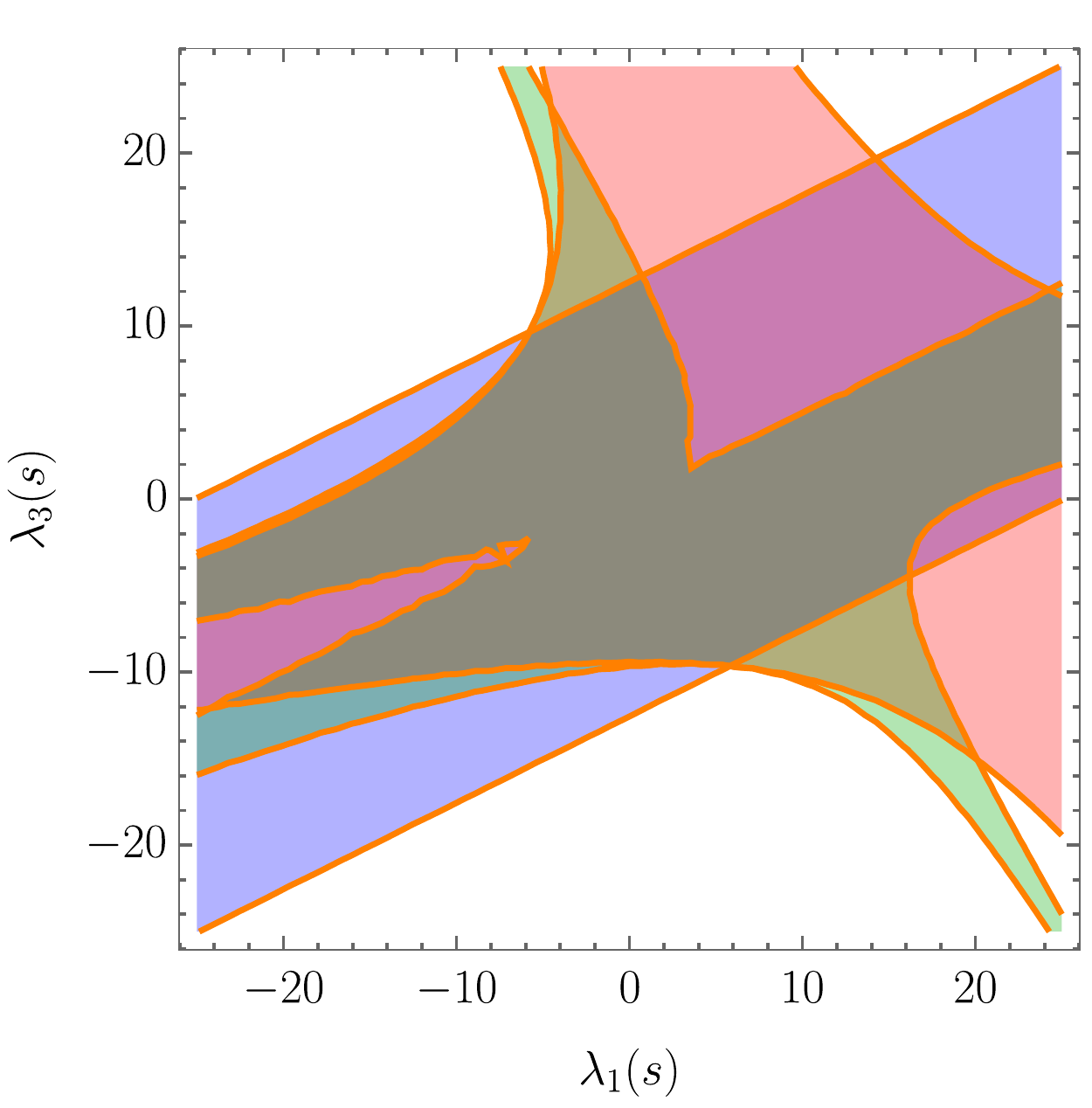}} 
 \caption{Viable parameter space for two individual eigenvalues based on the LO (blue), NLO (green), and NLO+ (red) perturbativity conditions, respectively.}
  \label{fig:U2i}
\end{figure}

Figure~\ref{fig:U2c} shows the bounds on $\lambda_1$ and $\lambda_3$ that result when the various perturbativity conditions,~\eqref{eq:Ubounds}, are applied. 
The parameter space shaded blue, orange, green, red, and purple is ruled out by perturbativity bounds from eigenvalues whose LO terms are proportional to $4 \lambda_1 + \lambda_3$, $- \lambda_1 + 2 \lambda_3$, $2 \lambda_1 - \lambda_3$, $\lambda_1$, and $\lambda_3$, respectively. 
There is a noticeable  difference in the viable parameter when the square of the NLO piece is included versus when it is not, as shown in panels~(b) and~(c) of Fig.~\ref{fig:U2c}. 
This is reminiscent of a result in Ref.~\cite{Falkowski:2016cxu}, where the bounds obtained on the coefficients of some dimension-6 operators in the SMEFT depend noticeably on whether or not the square of the dimension-6 amplitude is included in the calculation of the cross section under consideration. 
This parameter space could be further constrained theoretically by required the potential to be bounded from below, \textit{etc}.
Figure~\ref{fig:U2i} shows in blue, green, and red the viable parameter space for two individual eigenvalues based on the LO, NLO, and NLO+ perturbativity conditions, respectively. 
The contours in panels~(a) and~(b) of Fig.~\ref{fig:U2i} are colored blue and orange, respectively, to match the color coding of the eigenvalues in Fig.~\ref{fig:U2c}.
The parameter space viable at NLO (green) in Fig.~\ref{fig:U2i} is determined using both the upper and lower limits on $a_0$, which, in contrast, are illustrated separately in Fig.~\ref{fig:U2c} in panels~(b) and~(d), respectively.
The different perturbativity criteria yield different viable parameter spaces.
Perhaps the simplest way to test the convergence of $a_0$ as a perturbative series is to see which choices of parameters are considered viable by multiple perturbativity criteria.

%%%%%%%%%%%%%%%%%%%%%%
\section{Discussion}
\label{sec:sum}
In the section we discuss computing the leading order partial wave matrix and one-loop scalar beta function, and the validity of our approximation, before summarizing our algorithm for finding the NLO perturbativity bounds in a given theory.

One of the advantages of this approach is its simplicity in that it only relies on knowledge of the leading order partial wave matrix and one-loop scalar beta function.
In particular, the renormalization group equations in a general quantum field theory to two-loop order have been known for some time, and software exists to derive them in a specific model~\cite{Lyonnet:2013dna}. 
In the extreme energy limit $s \gg M_i^2$, which is sufficient to consider if one is only interested in bounding dimensionless couplings, the method of Ginzburg and Ivanov can be used to simply the computation of $\mathbf{a}_0^{(0)}$~\cite{Ginzburg:2005dt}. 
Though they originally considered the 2HDM, their argument can be used for any renormalizable theory where the matter fields have definite representations under the gauge group of theory. 
In particular at high energies masses and mixings are not important, and $\mathbf{a}_0^{(0)}$ will be block diagonal with blocks of definite representations under the gauge and global symmetries of the theory~\cite{Ginzburg:2005dt}. 
In gauge theories, scattering amplitudes involving longitudinally polarized vector bosons should be included when determining the bounds on the quartic couplings of the theory. 
Their inclusion can be greatly simplified through the use of the Goldstone Boson Equivalence Theorem, see \textit{e.g.}~\cite{Bagger:1989fc}. 
Lastly, we note that in gauge theories (with spontaneous symmetry breaking) there is the additional complication of preserving gauge invariance. 
In a mixed $\overline{\text{MS}}$/on-shell scheme, gauge independence can be spoiled unless tadpole diagrams are properly taken into account. 
This can be done by generalizing the SM results of Fleischer and Jegerlehner~\cite{Fleischer:1980ub} to the theory under consideration. 
In fact this has been done in the 2HDM~\cite{Krause:2016oke, Denner:2016etu, Krause:2016xku} and the SM Effective Field Theory (SMEFT)~\cite{Hartmann:2016pil}.

In considering how well our approximation does at capturing the exact NLO results, theories can be placed into one of three categories, broadly speaking.
The first class is theories are those without spontaneous symmetry breaking. 
Here there exists a renormalization scheme such that the approximation is actually exact.
This is simply because all of the counterterms for the quartic couplings are independent of the mass counterterms, and the finite parts of the quartic couplings can be chosen to cancel any potential wavefunction counterterm contribution.
The second class of theories is those with spontaneous symmetry breaking and a $\phi^4$-like scalar sector.
This class includes the SM and the 2HDM, theories for which this approximation is known to be good one~\cite{Grinstein:2015rtl, Cacchio:2016qyh}.
In multi-$\phi^4$ theory with spontaneous symmetry breaking the wavefunction renormalization is finite at one-loop, but more importantly it is parametrically identical to the 1PI one-loop contribution.
Thus, in $\phi^4$-like theories is that there is no qualitative difference between the different NLO contributions, and there is a qualitative difference between the LO and NLO contributions
This suggests that at the very least including some NLO contributions should give a good qualitative estimate of the full NLO contribution, which may be sufficient if one wishes to test the convergence of $a_0$ in perturbation theory through NLO.
Finally, the third class of theories are those with spontaneous symmetry breaking and scalar cubic interactions whose cubic couplings are not related to its quartic couplings.
In contrast with the first two classes of theories, it is not necessarily the case that this approximation will give a good description of the exact NLO result.
The reason being that the wavefunction renormalization contribution is parametrically different from both the LO and the 1PI NLO contributions to the partial wave amplitude.
On the other hand, if the cubic coupling of interest is much smaller than the internal masses in the wavefunction renormalization diagrams, the approximation may still be a good one as the theory is approaching the $\phi^4$ limit in this scenario.
Examples of theories of this type include extending the SM with a real scalar singlet without a $\mathbf{Z_2}$ symmetry, or with scalars that are color singlets and $SU(2)_L$ triplets. This includes the original Georgi-Machacek model, but not its generalizations~\cite{Logan:2015xpa}. Additionally, the 2HDM plus a pseudoscalar singlet falls into this class, see for example~\cite{Goncalves:2016iyg} and the references therein.
Another thing to keep in mind in our approach to finding the NLO perturbativity bounds is the quartic couplings entering into the partial-wave amplitudes are necessarily running couplings evaluated at an energy scale much larger than the other scales in the problem. 
A LO analysis is simpler in the sense that expressions could simply involve ordinary quartic couplings. 
One could always RG-improve the LO bounds by replacing the ordinary couplings with running couplings with no penalty.
This is the leading-log approximation. 
However, to do the opposite, replace running couplings in the NLO expressions with ordinary couplings, would be a further approximation.

To summarize, in this work we first derive a formula for the functional form of the partial wave matrix for high energy $2 \to 2$ scalar scattering in a general renormalizable theory.
This allows us to construct an algorithm for approximately computing the NLO perturbativity bounds on the scalar quartic couplings of a general renormalizable theory.
We expect the approximation to be a good estimate of the full NLO contribution in theories whose scalar sector is $\phi^4$-like. 
By this we mean theories where either there are no cubic scalar interactions, or the cubic couplings are related to the quartic couplings through spontaneous symmetry breaking.
The approximate NLO result only requires knowledge of the leading order matrix of partial wave amplitudes, and the one-loop scalar contribution to the beta function of each quartic coupling that is to be bounded, both of which are quantities that are relatively easy to determine.
The algorithm for finding the eigenvalues of the partial wave matrix at approximate next-to-leading order is as follows:
\begin{enumerate}
\item Given the leading order partial wave matrix, $\mathbf{a}_0^{(0)}$, find its eigenvalues, $a_0^{(0)}$, and eigenvectors, $\vec{x}_{(0)}$. 
This may need to be done numerically.
\item The $\beta$-function contribution to the NLO eigenvalues, $a_{0,\beta}^{(1)}$, is given by Eq.~\eqref{eq:EVper} with
\begin{equation}
\label{eq:Sbeta}
\mathbf{a}_{0,\beta}^{(1)} = - \frac{3}{2} \left. \mathbf{a}_0^{(0)}\right|_{\lambda_m \to \beta_{\lambda_m}} .
\end{equation}
\item Using the unitarity of the theory, the $\sigma$-contribution to the NLO eigenvalues is
\begin{equation}
\label{eq:Ssigma}
a_{0,\sigma}^{(1)} = \left(i - \frac{1}{\pi}\right) \left(a_0^{(0)}\right)^2 .
\end{equation}
\item The NLO contribution is given by the sum of the two pieces, $a_0^{(1)} = a_{0,\beta}^{(1)} + a_{0,\sigma}^{(1)}$. 
\end{enumerate}
In addition, Mathematica and Jupyter Notebook packages implementing this algorithm, $\mathtt{NLOUnitarityBounds}$, are available at~\url{https://github.com/christopher-w-murphy/NLOUnitarityBounds}.
The natural interpretation of the couplings appearing in the NLO partial wave amplitudes are RG improved couplings evaluated at a scale much larger than the typical scales of the theory.
Therefore these corrections are naturally useful in analyses at high scales. 
Once $a_0^{(1)}$ is known perturbativity bounds on quartic couplings can be obtained at either leading order or next-to-leading order in theories with $\phi^4$-like scalar sectors, enabling a test of the convergence of the $a_0$ as a perturbative series among other studies.
This is highlighted in the example we present.
Specifically, there is a noticeable difference in the viable parameter when the square of the NLO piece is included versus when it is not.

%%%%%%%%%%%%%%%%%%%%%%%%%%%%%%%%%%%%%%%%%%%%%%
\begin{acknowledgments}
We thank D. Chowdhury, H. Davoudiasl, S. Dawson, O. Eberhardt, B. Grinstein, and P. Uttayarat for useful discussions. This work was supported by the US DOE under grant contract DE-SC0012704.
\end{acknowledgments}

%%%%%%%%%%%%%%%%%%%%%%%%%%%%%%%%%%%%%%%%%%%
\providecommand{\href}[2]{#2}\begingroup\raggedright\endgroup

%%%%%%%%%%%%%%%%%%%%%%%%%%%%%%%%%%%%%%%%%%%%%
\end{document}